\documentclass[11pt,showpacs,preprintnumbers,aps,amsmath,amssymb]{revtex4}
\usepackage{latexsym}
\usepackage{graphics}
\usepackage{epsfig}
\newcommand{\be}{\begin{equation}}
\newcommand{\ee}{\end{equation}}
\newcommand{\bear}{\begin{eqnarray}}
\newcommand{\eear}{\end{eqnarray}}
\newcommand{\ba}{\begin{array}}
\newcommand{\ea}{\end{array}}

\newcommand{\CL}{{\cal L}}

\begin{document}
\title{Bounds on the Simplest Little Higgs Model Mass Spectrum Through Z Leptonic Decay}

\author{Alex G. Dias}
\affiliation{ {\small  Centro de Ci\^{e}ncias Naturais e Humanas, Universidade Federal do ABC \\
Rua Santa Ad\'{e}lia 166,  
Santo Andr\'{e}, SP 09210-170, Brazil.}}
\author{C. A. de S. Pires and P. S. Rodrigues da Silva}
\affiliation{ {\small  Departamento de F\'{\i}sica, Universidade Federal da
Para\'{\i}ba, Caixa Postal 5008, Jo\~ao Pessoa, PB 58051-970,
Brazil.}}

\date{\today}


\begin{abstract}
We derive the leptonic neutral current  in the simplest little Higgs model and compute the contribution of the model to the decay  width
$Z \rightarrow e^+e^-$. Using the precision electroweak data we obtain a  strong lower bound $f\geq 5.6$~TeV at $95\%$ C.L. on the characteristic energy scale of the model. It results in a lower bound for the new gauge bosons $W^{\prime\pm}$ and $Z^{\prime}$ as being $M_{W^{\prime\pm}}\geq 2.6$~TeV and $M_{Z^{\prime}}\geq 3.1$~TeV, respectively. We also present the allowed values of the $k=f_1/f_2$ which is the parameter relating the two vacuum expectation values of the scalar triplets in the model, and the $\mu$ parameter of a quadratic term, involving the triplets, necessary to provide an acceptable mass range for the standard Higgs boson.  
\end{abstract}

\maketitle




\section{Introduction} \setcounter{equation}{0}
\label{sec1}

The Little Higgs idea is a new purpose for solving the little  hierarchy problem that arises in gauge models for the electroweak interactions, where the symmetry breaking is triggered by the Higgs mechanism~\cite{idea}. The main idea behind Little Higgs is that  the Higgs field is  a pseudo Goldstone boson of an extended global symmetry. This global symmetry is violated by gauge and Yukawa interactions such that the Higgs field acquires  mass from one loop radiative corrections. But now  the quadratically divergent contributions are canceled out due to the existence of new particles at TeV scale. 

Among the realistic realizations of the mechanism behind the Little Higgs the most economical one is based on the $SU(3)_L\otimes U(1)_X$ gauge group. This model is known as the Simplest Little Higgs model (SLHM)~\cite{SLHM}. Besides recovering the standard particles spectrum, the SLHM predicts new vector gauge  bosons, three of them are neutral and one is charged. In the fermion sector, the model has three new quarks, one of them being a heavy top-like quark $T$. Finally, the scalar sector is composed of two scalar triplets in a non-linear sigma model realization. In this way, after spontaneous symmetry breaking, only the standard Higgs and a new pseudo-scalar survive.

 Thus, in fact, the physical scalars are pseudo Nambu-Goldstone bosons and their masses generated at one loop level are proportional to the  gauge and Yukawa coupling constants times the logarithm of the cutoff  $\Lambda\approx 4\pi f$, with $f$ the energy scale related to the global symmetry breakdown present in Little Higgs Models. The new particles in the SLHM have masses related to this scale, so that a lower bound on $f$ reflects as a lower bound on these masses. Once the experimental support to a specific Little Higgs Model does not concern only the Higgs production, but also the identification of a new particle content, such a lower bound for $f$ turns out to be an important issue. 

On the phenomenological side, for example, we have that atomic parity violation in Cesium requires $f\geq 1.7$ but the most  stringent constraints come from the oblique parameter $S$ due to the $Z^{\prime}$ which requires $f\geq 5.2$ TeV at $95\%$ C.L.~\cite{lhstwy}. Other important phenomenological constraints on Little Higgs models which are relevant for bounding the $f$ parameter are presented in Ref.~\cite{Kilian1} by considering a general analysis of electroweak precision data.

It would be interesting to pursue new possibilities of improving the bounds on $f$ and we wish to do this by looking at the $Z$ decay width, which is one of the best results from high energy precision experiments. However, if we are to compute the full $Z$ decay width,
we should have in mind that the SLHM is basically a kind of 3-3-1
model~\cite{331nr}, whose quark sector is plagued with mixing between ordinary quarks and the heavy new ones~\cite{Mix331}. 
Instead, in this work we are going to extract a constraint on the scale $f$ of SLHM from the $Z \rightarrow e^+e^-$ decay only. It is evident that this particular channel would be cleaner than computing the total decay width, meaning that no hadronic current enters in the computation, which turns our approach independent on several unknown parameters.

We are going to see that this choice turns out to be a little more restrictive than previous ones, pushing the lower bound on $f$ to about $5.6$~TeV. Also, by looking at the minimum of the scalar potential we are able to determine the range of values for the $\mu$ scale and the $k$ factor, which relates the two scalar triplets vacuum expectation values (VEV)s, $f_1$ and $f_2$, defining the acceptable range of values for the Higgs mass.

\section{The SLHM minimal content}
\label{sec2}
Here we resume the model used in this work~\cite{SLHM}. The model is constructed from a global  SU(3)$_1\times$SU(3)$_2$ with two scalar triplets: $\Phi_1$ transforming as a triplet of the factor  SU(3)$_1$ but as a singlet  of the factor SU(3)$_2$ and $\Phi_2$ transforming as a triplet of the factor  SU(3)$_2$ but as a singlet  of the factor SU(3)$_1$.  This global symmetry is spontaneously broken, at a scale $f$,  to $SU(2)_1\times SU(2)_2$ by the scalar VEVs. It results in ten Nambu-Goldsone bosons which could be described as a non-linear sigma model  SU(3)$_1\times$SU(3)$_2$/SU(2)$_1\times$SU(2)$_2$, if  the radial modes are assumed to get mass above the $f$ scale. Therefore,  when the diagonal subgroup $SU(3)_W\times U(1)_X$ of the global symmetry  is turned into a gauge symmetry, the scalar fields break the gauge group to the Standard Model (SM) $SU(2)_L\times U(1)_Y$ group at the scale $f$. Five of the ten degrees of freedom associated with the Nambu-Goldstone bosons are absorbed into the longitudinal components of the new gauge bosons, which turn them massive at this scale. The remaining degrees of freedom  in the scalar triplets stay massless at tree level but become massive as a result of the radiative corrections. This happens because introduction of the gauge and Yukawa interactions breaks explicitly the global SU(3)$_1\times$SU(3)$_2$ symmetry.  Thus, in fact, the physical scalars are pseudo Nambu-Goldsone bosons and their masses, generated at one loop level, are proportional to the  gauge and Yukawa coupling constants times the logarithm of the cutoff  $\Lambda$.

In the non-linear description,  the scalar triplets, which transform as  $\sim(3\,,\,-1/3)$ under $SU(3)_W\times U(1)_X$,  can be parameterized as 
\begin{eqnarray}
\Phi_1 =\mbox{exp}  \left \{ i \frac{ \Theta }{kf}\right \} 
\vec{ f_1}
\,\,\,, \,\,\,\,\,\,\,\,\,
\Phi_2 =\mbox{exp}\left \{ -i \frac{k\Theta }{f} \right \}
\vec{ f_2},
\label{scalartriplet}
\end{eqnarray}
Where we define,  $k=\frac{f_1}{f_2}$, the scale  $ f^2=f^2_1 + f^2_2$ and the vacuum direction is assumed $ \vec{ f_i}= (0 \,\,0\,\, f_i)^T$, with $T$ meaning transposition; $\Theta$ is the matrix 
\begin{eqnarray}
	\Theta=  \left[   \begin{array}{ccc}
 -\frac{\sqrt2}{4} \eta & 0 & h^0 \\
 0 & -\frac{\sqrt2}{4}\eta  & h^- \\
 h^{0 *}  & h^{+} & \frac{\sqrt2}{2}\eta 
\end{array}
 \right] \,\, , 
\label{nbmatrix}
\end{eqnarray}
\noindent
$h^0$  and $h^-$ form a doublet under  $SU(2)_L$, $h=\left( h^0\,\,,\,\, h^- \right)^T$, which is identified with the standard model Higgs doublet and $\eta$ is a pseudo-scalar singlet field.

The SLHM  have the  minimal lepton content composed according to the following representations of SU(3)$_L$ $\otimes$ U(1)$_X$ 
\begin{eqnarray}
& & \L_{aL} = \left ( \nu_a \, \, e_a \, \, \nu^C_a\right )^T_L \sim(3\,,\,-1/3)\,, 
\nonumber\\
& & e_{aR}\,\sim(1,-1),
\end{eqnarray}
where $a = 1,\,2,\, 3$. Right-handed singlet neutrinos can also be added but they are irrelevant for our present purposes.

The quarks can be chosen as composing  the following representations: $\Psi_{Q_{1}}= (d \,\, u\,\, D )^T_L$ and  $\Psi_{Q_{2}}=(s \,\, c\,\, S )^T_L$ both transforming as  $\sim({\bf3}^{*},0)$, and $\Psi_{Q_3}=( t \,\, b \,\, T )^T_L$ transforming as $\sim( {\bf 3}, 1/3)$; with the right-handed components $u_ R,\,\, c_R,\,\, t_R,\,\, T_R$ transforming as $\sim( {\bf 1},2/3)$; and $d_R,\,\, s_R,\,\, b_R,\,\,D_R,\,\, S_R$  transforming as $\sim( {\bf 1},-1/3)$.  The new heavy quarks are $D$, $S$, and $T$ which, according to the global symmetry of the little Higgs model, cancel out one loop quadratic divergences for the Higgs mass due to $d$, $s$ and $t$ quarks. This choice for the quark representations is special in the sense that  anomalies cancellation involves the three fermion generations altogether \cite{331nr}.  In fact, the dominant contribution coming from fermions for the effective potential is due to the top quark and its heavy partner in the third generation. This is because the known quarks are too light compared to the  top quark. With the scalar triplets in Eq.~(\ref{scalartriplet}) transforming as $ \Phi_1$,$\Phi_2\sim({\bf3},-1/3)$, the relevant Yukawa interactions to compute the effective potential are then
\bear 
\CL_q = \lambda_{t1} \overline{\Psi}_{Q_{3}} \Phi_1 t_R  +
\lambda_{t2} \overline{\Psi}_{Q_{3}} \Phi_2 T_R + h.c. 
\label{lq}
\eear 
The Yukawa couplings are taken to be diagonal, avoiding
crossed terms such as $\overline{\Psi}_{Q_{3}} \Phi_1 T_R $ in
order to simplify the analysis.

As we have said, the  little Higgs mechanism is such that the tree level interaction violating the global symmetry, like that in Eq.~(\ref{lq}), generates an effective potential for the Higgs field so that it has a  nontrivial VEV, $\langle h\rangle=\left( v\,\,,\,\, 0\right)^T$. Thus, the reduction of the SM group down to the electromagnetic group is dynamically induced following the Coleman-Weinberg mechanism \cite{colwein}.

As it happens, when a gauge symmetry larger than the SM one is assumed, and supposing that it is spontaneously broken at the scale $f > v$, the couplings of the fields representing  the known gauge bosons, $Z$ and $W^\pm$, deviate by terms proportional  to $v^2/f^2$ from those predicted at the tree level by the  SM~\cite{com1}.  

Our aim is to extract the leptonic $Z$ boson couplings up to order $v^2/f^2$ to constrain this ratio through Z decay into leptons. The Z decay into neutrinos is not so restrictive as the decay  in charged leptons.  The reason is that  the axial-vector neutrino coupling receives a $v^2/f^2$ correction which is suppressed by other factors and, more importantly, the invisible width $\Gamma_{Z\rightarrow\nu\nu}$ is not as precise as $\Gamma_{Z\rightarrow l^+l^-}$.  

To obtain the masses and the eigenstates of the gauge bosons the  following parameterization for  the VEVs is used:
\begin{eqnarray}
& & \langle \Phi_1 \rangle=\frac{kf}{\sqrt{1+k^2}} \left [
i \sin \left ( \frac{ v}{kf} \right ) \,\,\,0 \,\,\,\cos \left ( \frac{v}{kf} \right ) \right ] ^T
\,, \nonumber\\
 \nonumber\\
& & \langle \Phi_2 \rangle  = \frac{f}{\sqrt{1+k^2}} \left [
-i  \sin  \left ( \frac{kv}{f} \right ) \,\,\,0 \,\,\,   \cos \left (  \frac{kv}{f} \right ) \right ] ^T, 
\label{VEV}
\end{eqnarray}
with $v$  being the  VEV of the neutral scalar $h^0$.

 The gauge boson masses are obtained from the kinetic terms for the scalar fields
$\Phi_i$,
\begin{eqnarray}
{\cal L}=|\left(\partial_\mu-\frac{i}{2}gW^a_\mu T^a+
\frac{i}{3}g_X B_\mu  \right)\Phi_i|^2, 
\label{kineticterm}
\end{eqnarray}
where $T^a$ represents a generator of the gauge group $SU(3)_W$,  
with $a=1...8$; $B_\mu$  is the gauge boson associated to the abelian 
gauge group $U(1)_X$.

Among the  neutral vector bosons, besides the photon $A_\mu$, there will be three massive states called $Z_{1 \mu}$, $Z_{2 \mu}$ and $U^0_\mu$, which are linear combinations of  the symmetry fields $W_{3 \mu}$, $W_{8 \mu}$, $B_\mu$ and $W_{5 \mu}$, and a fourth massive state which decouples from the other states and is identical to its own symmetry eigenstate $W_{4\mu}$. The structure of the VEVs required by the little Higgs mechanism is clearly distinct, concerning the gauge bosons, compared with some well known $SU(3)_W\times U(1)_X$ models. In the last ones it is still possible to have two real and a non-hermitian  massive neutral fields, according to the chosen direction of the VEVs. The bilinears involving the gauge fields from which the  neutral vector bosons are originated, taking into account  Eq.~(\ref{VEV}),  are 
\bear  
\CL^{^{mass}}_{neut.} & = & \left( \ba{cccc} W_3 & W_8 & B & W_5 \ea \right )
 \frac{{\cal M}^2_{neu}}{2}\left(
\ba{c}
W_3  \\
W_8 \\
B \\
W_5 \ea \right ) + \frac{g^2}{4}f^2 W_4^\mu W_{4\mu} \label{bneu}
\eear
Their mass matrix, which can be diagonalized exactly, is
\be
  {\cal M}^2_{neu}  = \frac{g^2}{2}\left( \ba{cccc}
a & \frac{1}{\sqrt3}a & -\frac{2b}{3}a & -u \\
\frac{1}{\sqrt3}a & -a+\frac{4}{3}f^2 & -\frac{2b}{3\sqrt3}(3a-2f^2) & 
\frac{1}{\sqrt3}u \\
-\frac{2b}{3}a & -\frac{2b}{3\sqrt3}(3a-2f^2) & \frac{4b^2}{9}f^2 & 
\frac{4b}{3}u \\
-u & \frac{1}{\sqrt3}u & \frac{4b}{3}u & f^2 \\
 \ea \right ) \label{mmz1z2}
\ee
with the definitions~\cite{com2} $t=g^\prime/g = \tan(\theta_W)$, 
\bear
 a &=& \frac{k^2 \sin^2 \left ( \frac{ v}{kf} \right ) + \sin^2 \left ( \frac{ kv}{f} \right ) }{1+k^2}f^2
\label{pa}\\
\nonumber\\
b^2 &=& \frac{g_X^2}{g^2}=\frac{3t^2}{3-t^2}
\label{bdef} \\ \nonumber \\
u &=&  \frac{k^2 \sin \left ( \frac{ v}{kf} \right ) \cos \left ( \frac{ v}{kf} \right ) - \sin \left ( \frac{ kv}{f} \right ) \cos \left ( \frac{ kv}{f} \right )}{1+k^2}f^2
\label{pu}
 \eear

We get the vector boson masses through diagonalization of Eq.~(\ref{mmz1z2}),
\bear
M^2_{Z_1} & = & \frac{1}{2} \left[
M^2_{Z^\prime} -\sqrt{M^4_{Z^\prime} - 4\delta_Z} \right]~, \,\,\,\,\,\,\,\,
M^2_{Z_2} = M^2_{Z^\prime} -  M^2_{Z_1} 
\label{AMZ12} \\
\nonumber\\
M^2_{U^0} &=& \frac{1}{2}g^2 f^2 \label{AMU0}~, \,\,\,\,\,\,\,\, 
M^2_{\gamma}  =  0,
\eear
where
 \bear
 \delta_Z &=& g^4\, \frac{(1+t^2)}{(3-t^2)} 
 \frac{k^2 f^4}{(1+k^2)^2} \sin^2 \left ( \frac{(1+k^2) v}{kf} \right ) 
\label{dz}\\
\nonumber\\
M^2_{Z^\prime} &=& \frac{2}{3-t^2}\,g^2f^2, 
\label{MZ}
 \eear
 Now we wish to raise a point that has  been first discussed in this model in Ref.~\cite{lhued}.  We observe that these eigenvalues are invariant by the change $k \rightarrow 1/k$. This symmetry is nothing more than the fact that the gauge interactions do not distinguish one scalar triplet from the other, i. e., there is no physical effect in performing the change $f_1\leftrightarrow f_2$. In fact, the effective potential exhibits such a symmetry once it depends only on  $ \vert\Phi_1\vert^2\vert\Phi_2\vert^2-\vert\Phi_1^\dagger\Phi_2\vert^2$. Another point which should be mentioned concerns this invariance in the limit $k\rightarrow 0$ (or equivalently $k\rightarrow \infty$)  keeping the ratio $v/f$ fixed. It leads to  $\delta_Z\rightarrow 0$ so that spontaneous symmetry breaking of $SU(2)_W\times U(1)_Y$ does not occur in this limit, once the Higgs potential goes, in this case, to zero. Thus, the particle masses show a dependence on the $k$ parameter and these masses can be sensitive to this parameter even when the ratio $v/f$ is small. Actually, this is an  interesting fact since masses at the electroweak scale cannot be free from the high energy physics as we expect, if all low energy parameters originate from a previous  structure of a more fundamental underlying theory.

The neutral vector bosons eigenstates, in the basis $W_3$, $W_8$, $B$ and $W_5$, are  

\begin{eqnarray}
	Z^1_\mu= \frac{1}{N_{z1}}\left[ c_3 W^3_\mu +c_8 W^8_\mu +c_B B_\mu +c_5W^5_\mu \right] ,
	\label{z1}
\end{eqnarray}
where
\begin{eqnarray}
& & c_3 =2a(3+4t^2)-4(t^2+3)+9m_{z1}^2,\nonumber \\
& & c_8=  \sqrt3 \left[ 2a(3+4t^2)-4t^2 -3m_{z1}^2\right],
\nonumber \\ 	
& & c_B = 12t (1-m_{z1}^2),
\nonumber \\ 
& & c_5 = -  4u (3+4t^2). 
\label{coefsc}
\end{eqnarray}
With the normalization factor $N_{z1}=\sqrt{c_3^2+c_8^2+c_B^2+c_5^2}$ and $m_{z1}^2=2M_{Z_1}^2/g^2f^2$. 

The expansion of $Z^1$ above is sufficient for our purposes, and we can do the same for the other mass eigenstates,
 \begin{eqnarray}
      Z^2_\mu= \frac{1}{N_{z2}}\left[ d_3 W^3_\mu +d_8 W^8_\mu +d_B B_\mu +d_5W^5_\mu \right] ,
\label{z2vector}
\end{eqnarray}
where
\begin{eqnarray}
& & d_3 =2a(3+4t^2)-4(t^2+3)+9m_{z2}^2,\nonumber \\
& & d_8=  \sqrt3 \left[ 2a(3+4t^2)-4t^2 -3m_{z2}^2\right],
\nonumber \\ 	
& & d_B = 12t (1-m_{z2}^2),
\nonumber \\ 
& & d_5 = -  4u (3+4t^2).
\label{coeficientz2}
\end{eqnarray}
For the third neutral massive vector boson and the photon we have 
 \begin{eqnarray}
      U^0_\mu= \frac{1}{\sqrt{4u^2+(2a-1)^2}}\left[  uW^3_\mu +\sqrt3 u W^8_\mu +(2a-1)W^5_\mu \right] 
\label{uvector}
\end{eqnarray}
and 
\begin{eqnarray}
	A_\mu= \frac{1}{\sqrt{3+4t^2}}\left[  \sqrt3 t W^3_\mu -tW^8_\mu + \sqrt3 B_\mu \right],
\label{photonvector}
\end{eqnarray}
respectively.

We can then extract the vector and axial-vector couplings of neutral gauge bosons with leptons, $g_V$ and $g_A$, by considering the above eigenstates and the associated interaction Lagrangian. Since we are interested on leptonic $Z$ decay, we concentrate on $Z^1$ couplings with the charged leptons only,

\begin{eqnarray}
{\mathcal  L}^{NC}&=& - \frac{g}{2 c_W} 
\sum_{l}\overline{l}\gamma^\mu(g^{l}_{_V}-g^{l}_{_A}\gamma^5)l  Z_{1\mu}. 
\label{lncz}
\end{eqnarray}
Where $l= e, \mu, \tau, $ stands for the charged leptons electron, muon and tau, respectively, and the vector and axial-vector couplings are the following 

\begin{eqnarray}
&& g^{l}_{_V}= \frac{2c_W }{N_{z1}}(3-8t^2)(1-m^2_{z1}) 
\approx -\left( \frac{1}{2}-2s^2_W\right)\left(1-\frac{(1-4c^2_W)}{8c^4_W}\frac{v^2}{f^2}\right)
\nonumber \\ 
\nonumber \\
&& g^{l}_{_A}= \frac{2c_W }{N_{z1}}(3+t^2)(1-m^2_{z1}) 
\approx -\frac{1}{2}+\frac{(1-4c^2_W)}{16c^4_W}\frac{v^2}{f^2}. 
\label{gAgVnus}
\end{eqnarray}
We will use the expansion in powers of $v^2/f^2<<1$ till first order to constrain the values of $f$  according to the precision measurement of the width, $\Gamma_{Z\rightarrow l^+l^-}$.

\section{The leptonic decay  $Z^1 \rightarrow e^+ + e^-$}

The partial width for a leptonic decay 
$Z^1 \rightarrow l^+ + l^-$ is given by~\cite{pdg},
\begin{eqnarray}
	\Gamma_{Z^1 \rightarrow l^+l^-}=\frac{G_F}{6\sqrt{2}\pi}m^3_{Z^1}
\left[ ({\bar g}^l_V)^2+({\bar g}^l_A)^2 \right]
\times(1+\delta\rho+\delta\rho_l+\delta_{\mbox{QED}}).
	\label{zdecay}
\end{eqnarray}
In this expression we should have in mind that the vector and axial-vector $Z-l-\bar{l}$ couplings, ${\bar g}^l_V$ and ${\bar g}^l_A$, respectively, comprise one-loop and higher order electroweak and internal QCD corrections~\cite{pdg,zfitter}, through the form factors $\delta\rho_l$ and $\kappa_l$. 
The parameter $\kappa_l$ embodies the radiative corrections that modify the Weinberg mixing angle, yielding the so called effective angle, $s_{W(eff)}^2\equiv \kappa_l s_W^2$. Then, in Eq.~(\ref{zdecay}), the vector and axial-vector couplings ${\bar g}^l_V$ and ${\bar g}^l_A$ are given by the same expression as in Eq.~(\ref{gAgVnus}) but with  $s_W^2$ replaced by $s_{W(eff)}^2$. 
The term $\delta\rho$ is the deviation from SM prediction for the $\rho$ parameter, $\rho\equiv M_Z^1 c_W/M_W = 1+\delta\rho$,  taking into account contributions of the gauge group structure of SLHM only. Considering Eq.~(\ref{AMZ12}) and the $W$ quadratic mass, at the required order, given by
\begin{eqnarray}
 M^2_{W} = \frac{g^2v^2}{2}\left[1-
\frac{1}{3} \frac{v^2f^2}{f_1^2f_2^2} + \frac{m^2_{W}}{M^2_{Ch}}
\right]
\label{mw}
\end{eqnarray}
the SLHM contribution to $\delta\rho$ reads,
\begin{eqnarray}
\delta \rho\approx\frac{v^2}{8f^2}(1-tg_W^2)^2\,.
\label{drho}
\end{eqnarray}
Also, $\delta_{\mbox{QED}}$ accounts for the final state photon radiation,
\begin{equation}
	\delta_{\mbox{QED}}=\frac{3\alpha(s)}{4\pi}Q^2\,,
	\label{QEDcorrect}
\end{equation}
where $\alpha$ is the QED coupling computed at the energy scale $s$, 
while $Q$ is the lepton charge.

In order to obtain a prediction for the standard model partial $Z$ decay width into $e^+e^-$ we take the input parameters~\cite{lepewwg}, 
$M_{Z^1}=91.1875$~GeV, $G_\mu=1.16637 10^{-5}$~GeV$^{-2}$, 
$\alpha(M_{Z^1})=1/128.95$, $m_{top}=175$~GeV, $M_H=150$~GeV 
and $s_W^2=0.22335$. These parameters can be used to obtain the form 
factors for the decay $Z\rightarrow e^+e^-$ using the Zfitter package
~\cite{zfitter}, yielding
\begin{equation}
	\delta\rho_{e}=0.00531
\end{equation}
and
\begin{equation}
	s_{w(eff)}^2=0.2315
\end{equation}
which translates into $\kappa_e =1.0367$. Plugging these parameters into Eq.~(\ref{zdecay}) together with the limit $f\rightarrow\infty$ and $\delta\rho\rightarrow 0$, we obtain the standard model prediction
$\Gamma (Z\rightarrow e^+e^-)=83.99~\mbox{MeV}$. This value perfectly reproduces the fit reported in the Particle Data Group (PDG)~\cite{pdg}.

However, the experimentally observed partial decay width for this channel, according to LEP-II results is $\Gamma (Z\rightarrow e^+e^-)=83.91\pm 0.12~\mbox{MeV}$~\cite{pdg}, leaving much room for new physics. We can take advantage of this window in order to get a better bound on the $f$ scale in SLHM.

Now let us obtain the constraint over $f$. The new contributions to the $Z_1$ decay width concern the modifications on the vector and axial-vector couplings, presented in Eq.~(\ref{gAgVnus}) for a finite $f$, as well as the SLHM contribution to the $\rho$ parameter given by Eq.~(\ref{drho}). Considering the experimental result within 95$\%$ of C.L., we are able to obtain the lower bound 
\begin{eqnarray}
	f\geq 5.6\mbox{~TeV},
	\label{contraint}
\end{eqnarray}
from the highest value $\Gamma (Z\rightarrow e^+e^-)=84.145~\mbox{MeV}$. This bound is stronger than the one obtained solely from the $\rho$ parameter~\cite{SLHM}, and represents a small improvement with respect to that obtained through oblique corrections~\cite{lhstwy}. 

The effects of this constraint in the  gauge boson sector  translate in lower bounds on the masses of the new gauge bosons. Keeping only the dominant terms in Eqs.~(\ref{AMZ12}) and (\ref{AMU0}) 
\begin{eqnarray}
  M_{Z_2}\geq 3160 \,\,\mbox{GeV}	\,\,\,\,,\,\,\,\,\mbox{and}\,\,\, M_{U^0}\geq 2600 \,\,\mbox{GeV}.
	\label{ZWUmass}
\end{eqnarray}

There is also a new charged gauge boson, $W^{\prime}$, whose mass at first order is such that $M_{W^{\prime}}\approx M_{U^0}=gf/\sqrt{2} $. Therefore, we can conclude that production of new gauge bosons in this model will not be possible even at the LHC, where it is expected an energy around 1 TeV for each quark carried by the protons. 

Another distinguished particle in this model is the exotic quark,   $T$, partner of the top quark. It was observed that the minimum value for the $T$ quark mass would be~\cite{SLHM} 

\bear
 M_{T} &=& 2\sqrt 2 \frac{m_{t}}{v} \frac{kf}{1+k^2} \approx \frac{11k}{1+k^2}  \,\,\mbox{TeV},  
\label{MTmim}
\eear
with the top quark mass taken as $m_t\approx 175$ GeV. A rough estimate of energy available for particle production at LHC is that each quark in the $pp$ reaction carries one TeV. Therefore, $M_T$ could be inside the potential available energy for production in LHC if $k\leq 0.2$ or, equivalently, $k\geq 5$ which corresponds to a mass $M_{T}\leq 2$ TeV. The prospects for observing the $T$ quark at forthcoming colliders as well as the singlet scalar, characteristic of Little Higgs models, is analyzed in Refs.~\cite{cheung2006}, \cite{Kilian2}, \cite{Kilian3}.

\section{Higgs mass}

In this model the Higgs is a pseudo-Goldstone boson related to the spontaneous breaking of the $SU(3)_1 \times SU(3)_2$ global symmetry, which is also explicitly broken, resulting in a mass for the Higgs boson.  However in order to generate an acceptable value for the scale $f$, we have to include a scale in the potential through the term ``$\mu^2 \Phi_1^{\dagger}\Phi_2$''. Gauge and Yukawa interactions break the global symmetry and then provide a potential for the standard Higgs boson. At one-loop, and keeping only dominant terms involving the Higgs field to the fourth power, this potential is the following \cite{SLHM}, 

\begin{eqnarray}
V_h =\left(  \frac{\mu^2 (1+k^2)}{k} +   m^2 \right)h^{\dagger}h 
+\left( - \frac{1}{12}\frac{\mu^2 k^3}{f^2(1+k^2)^3} 
+  \lambda  \right)(h^{\dagger}h)^2  
\,,
\label{higgspotential}	
\end{eqnarray}
with 

\bear
  m^2 = - \frac{3}{16\pi^2 } 
 \left [ 2\lambda_t^2\,M_T^2 \,\ln\left(\frac{\Lambda^2}{M_T^2 }\right)
 -\frac{g^2}{4}M_{Z'}^2\, \ln\left(\frac{\Lambda^2}{M_{Z^\prime}^2}\right)
 -\frac{g^2}{2}M_{W^{\prime}}^2\,\ln\left(\frac{\Lambda^2}{M_{W^{\prime}}^2 }\right)
  \right ] , 
\label{m2}
\eear

\bear
  \lambda  &= & - \frac{m^2 (1+k^2)^2}{3f^2k^2}   - \frac{3}{64\pi^2 v^4}
 \left [4m_t^4\left (\frac{1}{2}+ \ln \frac{m_t^2}{M_T^2 } \right )
 -m_{Z}^4\left ( \frac{1}{2} + \ln \frac{m_{Z}^2}{M_{Z^\prime}^2} \right)
   \right.  \nonumber\\ \nonumber\\
  & & \left. \,\,\,\,\,\,\,\,\,\,\,\,\,\,\,\,\,\,\,\,\,\,\,\,\,\,\,\,\,\,\,\,\,\,\,\,\,\,\,\,\,\,\,\,\,\,\,\,\,\,\,\,\,
 -2m_W^4\left (\frac{1}{2} + \ln \frac{m_W^2}{M_{W^{\prime}}^2} \right)
  \right ], 
\label{lamb}
 \eear
$\lambda_t$ and $g$ are the top quark Yukawa and SU(2) gauge coupling constants, respectively.

\begin{figure}
\centering
\epsfig{file=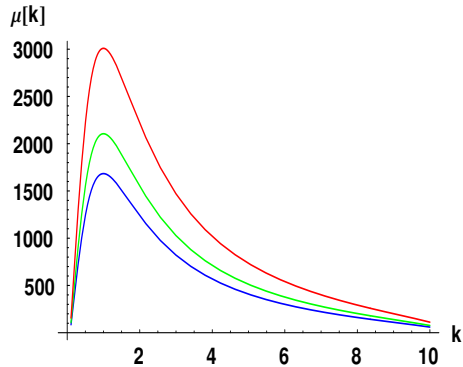,width=7cm,height=5.5cm,angle=0}
\caption{The $\mu$ parameter as a function of $k$. Starting from below, the curves are for $f$ = 5.6, 7 and 10 TeV, respectively. The symmetry $k\rightarrow 1/k$ is manifest.}
\label{fig:mu2}
\end{figure}

Once the scale $f$ is constrained according to Eq.~(\ref{contraint}) the physical condition that the true vacuum lies on the minimum of the potential leads to a relation involving $\mu$ and $k$, which is 

\begin{equation}
	\mu^2 = \frac{\left( v^2\lambda +  m^2 \right)}
{\left( \frac{v^2(1+k^2)^3}{12 k^3 f^2} - \frac{(1+k^2)}{k}\right)}\,.
	\label{muscale}
\end{equation}
Observe that in this equation, $\lambda$ and $m^2$ depend only 
on two free parameters, $f$ and $k$ (see Eq.~(\ref{m2}) and Eq.~(\ref{lamb}) above). 
We plot $\mu$ as a function of $k$ for the relevant range $5.6$~TeV $< f < 10$~TeV as shown in Fig.~\ref{fig:mu2}.
From this figure we see that $k=1$ yields the highest values for $\mu$, starting from 1.5~TeV. For a soft natural breaking of the global symmetry we would expect $\mu \approx 250$~GeV, a value that can be obtained for $k\leq 0.14$ or, equivalently, $k\geq 7$.

\begin{figure}
\centering 
\epsfig{file=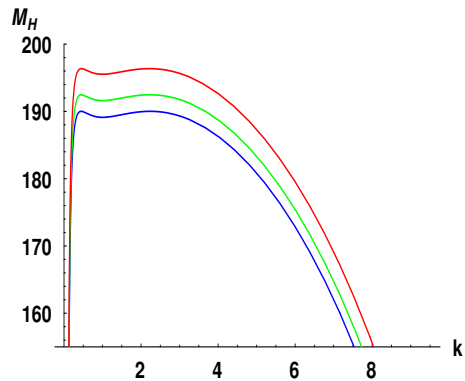,width=7cm,height=5.5cm,angle=0}
\caption{Higgs mass as a function of $k$. Starting from below, the curves are for $f$ = 5.6, 7 and 10 TeV, respectively. The symmetry $k\rightarrow 1/k$ is manifest. }
\label{fig:MassaHiggs}
\end{figure}

The Higgs mass derived from Eq.~(\ref{higgspotential})  is given by 

\begin{eqnarray}
m_h = 2v\sqrt{  \lambda - \frac{1}{12}\frac{\mu^2 k^3}{f^2(1+k^2)^3} }\,. 
\label{massahiggs}
\end{eqnarray} 
This Higgs mass is plotted against $k$ in Fig.~\ref{fig:MassaHiggs} and, as before, we have included three sample curves for $f=5.6$, $7$ and $10$~TeV. Observe that for an interval surrounding  $k=2$  there is no significant change in $m_h(k)$, even for values of $f$ higher than the bound derived in Eq.~(\ref{contraint}). Considering the LEP limit $m_h\geq 114$ GeV the restriction $0.105 \leq k \leq 9.5$ is obtained. Large values for $f$ require higher values for $\mu$ in order to satisfy the constraint in Eq.~(\ref{muscale}). This is a fine tunning situation which is not so severe as in the SM where, apparently, an adjustment has to be done to suppress a scale, $\Lambda_{_{cutoff}}\approx M_{Pl}$, near the Planck mass until the electroweak scale. Here we have $\Lambda\approx 70$ TeV.

Notice also that, from Fig.~\ref{fig:MassaHiggs}, even for a scale $f=10$~GeV the Higgs mass would be such that $m_h(k)< 200$~GeV, which settles the maximum Higgs mass on the appropriate scale to be reached by LHC, whatever the value of $f$ in the considered range. On the other hand the exotic $T$ quark could be produced at LHC only for a small range of the $k$ parameter, i. e.  $0.105 \leq k \leq 0.2$, or for $k\geq 5$.

\section{Conclusions}

We show here that the $Z^0$ width for charged leptons requires $f\geq 5.6$~TeV, which represents an improvement on the $f$ bounds~\cite{lhstwy}. Moreover, it implies a mass spectrum for the new gauge bosons in SLHM, $Z_2$, $U^0$ and $W^\prime$, that is mostly out of the LHC capability for their direct detection. Nevertheless, if the $k$ parameter is in the appropriate range, although very restricted as mentioned above, the new heavy quark $T$ can be produced at LHC and becomes a genuine signal of this SLHM. But once the new gauge bosons of this model cannot be produced in the present machines, only the production of $T$ quark is not sufficient to distinguish the SLHM from its viable competitors (see 
Ref.~\cite{Kilian3}  for a proposal to discriminate between Little Higgs models 
based in product group and simple group). In fact, there are two additional exotic quarks, $D$ and $S$, which cancel the one loop quadratic divergences in Higgs mass due to $d$ and $s$ quarks. Considering also the hypothesis  of producing  $D$ and $S$ exotic quarks, more work would be needed to distinguish the model we treat here from the known 3-3-1 models. 

It is fair to say that the option for using the Little Higgs idea for solving the (little) hierarchy problem could be criticized from a naturalness point of view. Indeed, there are studies concerning the amount of fine-tunning required by some known Little Higgs models~\cite{natlh}. In ref.~\cite{natlh} the authors conclude that, for scales $f\simeq 1$ TeV, the SLHM could be made the best behaved model under the fine-tunning analysis, but with a restricted region in the parameter space, where it is competitive with supersymmetric models.  An analysis for fine-tunning involving all parameters and  taking into account the constraint we present here for the scale $f$, would be necessary.  This would give us the regions in the parameter space where SLHM could be considered viable under this point of view.


\bigskip

{\bf Acknowledgments:}
The authors acknowledge the support of the State of S\~{a}o Paulo
Research Foundation (FAPESP), A. G. D., and  the Brazilian  National Counsel for Technological and Scientific Development (CNPq), C. A. S. P. and P. S. R. S.



\begin{thebibliography}{99}
\bibitem{com1} There are exceptions to this fact as for example the 3-3-1 models in Ref.~\cite{331np}, where a solution for VEVs can be found such that there is no deviation of the predicted SM couplings.
\bibitem{com2} Hereafter we will use the short definition for the Weinberg mixing angle, $s_W\equiv \sin \theta_W$ and $c_W\equiv \cos \theta_W$.
\end{thebibliography}
\end{document}